\documentclass[a4paper,10pt,twocolumn,twoside,journal]{article}

\usepackage{geometry}
\advance\oddsidemargin by -1in
\advance\evensidemargin by -1in
\advance\headsep by 2.8mm
\usepackage{cite}
\usepackage{hyperref}
\usepackage{url}
\usepackage{authblk}
\usepackage{mathptmx}
\usepackage{braket}
\usepackage{booktabs}
\usepackage{amsthm}
\usepackage{amssymb,amsmath,mdwlist,mathtools}
\usepackage{multirow}
\usepackage{multicol}
\usepackage{hhline}
\usepackage{algorithm}
\usepackage{algorithmicx}
\usepackage{algpseudocode}
\usepackage{graphicx}
\usepackage{caption}
\usepackage{subcaption}
\usepackage{adjustbox}
\usepackage{textcomp}
\usepackage{wrapfig}
\def\BibTeX{{\rm B\kern-.05em{\sc i\kern-.025em b}\kern-.08em
    T\kern-.1667em\lower.7ex\hbox{E}\kern-.125emX}}

\usepackage[normalem]{ulem}

\providecommand{\keywords}[1]{\textbf{\textit{Index terms---}} #1}

\title{Evaluation of Quantum and Hybrid Solvers\\ for Combinatorial Optimization}

\author[1]{Amedeo~Bertuzzi}
\author[1]{Davide~Ferrari}
\author[2]{Antonio~Manzalini}
\author[1,*]{Michele~Amoretti}
\affil[1]{\small Quantum Software Laboratory, University of Parma, Parma, Italy}
\affil[2]{\small TIM - Telecom Italia, Turin, Italy}
\affil[*]{\small Corresponding author: Michele Amoretti, michele.amoretti@unipr.it}

\date{}

\begin{document}

\maketitle

\thispagestyle{empty}
\pagestyle{empty}

\begin{abstract}
Academic and industrial sectors have been engaged in a fierce competition to develop quantum technologies, fueled by the explosive advancements in quantum hardware. While universal quantum computers have been shown to support up to hundreds of qubits, the scale of quantum annealers has reached three orders of magnitude (i.e., thousands of qubits). Therefore, quantum algorithms are becoming increasingly popular in a variety of fields, with optimization being one of the most prominent.
This work aims to explore the topic of quantum optimization by comprehensively evaluating the technologies provided by D-Wave Systems. To do so, a model for the energy optimization of data centers is proposed as a benchmark. D-Wave quantum and hybrid solvers are compared, in order to identify the most suitable one for the considered application. To highlight its advantageous performance capabilities and associated solving potential, the selected D-Wave hybrid solver is then contrasted with CPLEX, a highly efficient classical solver.
\end{abstract}


\section{Introduction} 
The advent of quantum technologies has opened new horizons in various information and communication technology fields, sparking unprecedented interest in the potential of quantum machines compared to classical computers. The use of quantum principles promises to overcome the intrinsic limitations of classical devices in solving problems previously considered impossible or excessively costly, ushering in a new era of computing capabilities.

Despite quantum technologies are still in their early stages, and undoubtedly have a long evolutionary path ahead, their performance capabilities are already visible, not only at a theoretical level but also in practical applications. Some quantum algorithms for carefully selected tasks require exponentially fewer computational steps than the best known classical algorithms. Optimization is frequently identified as a prime candidate to profit from such a revolution \cite{Abbas2023}.

This work aims to explore the topic of quantum optimization by comprehensively evaluating the technologies provided by D-Wave Systems~\cite{dwavesys}. To do so, a model for the energy optimization of data centers is proposed as a benchmark, because of its challenging complexity. D-Wave quantum and hybrid solvers are compared, in order to identify the most suitable one for the proposed benchmark. To highlight its advantageous performance capabilities and associated solving potential, the selected hybrid solver is then contrasted with CPLEX, a highly efficient classical solver.

The paper provides the following constributions:
\begin{itemize}
\item A Constrained Quadratic Model (CQM) concerning the energy optimization of a data center consisting of servers connected via switches in a complex tree topology. The challenging complexity of the model makes it suitable as a benchmark for quantum and hybrid solvers.
\item An overview of D-Wave's quantum and hybrid solvers, analyzing their capabilities and suitability for different classes of combinatorial optimization problems. Based on this analysis, the most suitable solver for the benchmark model is selected.
\item The implementation and experimental evaluation of a solution to the considered benchmark model based on Leap's hybrid CQM solver. In particular, a thorough comparative evaluation of the proposed solution with a classical CPLEX~\cite{ilogcplex} implementation is presented.
\end{itemize}

The rest of the paper is organized as follows. In Section~\ref{sec:prelimnary}, preliminary concepts on quantum optimization and suitable quantum computation models are introduced. In Section~\ref{sec:problem}, the considered optimization problem and the proposed CQM model are illustrated. In Section~\ref{sec:solvers}, the quantum and hybrid solvers provided by the D-Wave platform are presented in detail. In Section~\ref{sec:results}, the results of the experimental evaluation are presented and discussed. The paper is concluded with a discussion of the lesson learned and an outline for future work.

\section{Preliminary Concepts} 
\label{sec:prelimnary}

\subsection{Combinatorial Optimization Problems}
\label{sec:optimization}
The field of mathematical optimization plays a key role in solving several practical problems, ranging from intelligent resource allocation to complex strategic decision-making, applicable to an infinite variety of industrial and logistical fields. The theory of optimization aims to provide tools and methodologies for finding an optimal or near-optimal solution within the set of possible ones, given a well formulated problem.

Optimization problems are mainly divided into two classes based on the involved variables: if continuous, they are referred to as continuous optimization problems; if discrete, they are termed combinatorial optimization problems. This work delves into the latter category.

A combinatorial optimization problem is defined by a mathematical model that describes constraints and ways to compute solutions. A mathematical model consists of three main structures:
\begin{itemize}
    \item \textbf{Sets of Variables}: These represent the variables of the optimization problem, each belonging to a mathematical numerical set (binary, Ising, positive integers, etc.);
    \item \textbf{Constraints}: These represent the limitations on solution sets, modeling real constraints related, for example, to the mutual exclusivity of certain elements;
    \item \textbf{Objective Function}: It is a means to maximize or minimize a numeric value of interest.  Based on the values of the variables in the solution set and the parameters associated with them, the objective function calculates an objective value used to evaluate the solution set and compare it with others.
\end{itemize}

Mathematical models further divide based on the presence of constraints and the type of variables they accept. This work primarily focuses on the following types:
\begin{itemize}
    \item \textbf{CQM} (Constrained Quadratic Model): Models with constraints that use boolean and integer variables;
    \item \textbf{BQM} (Binary Quadratic Model): Models without constraints that use only boolean variables;
    \item \textbf{DQM} (Discrete Quadratic Model): Models without constraints that use only discrete variables;
    \item \textbf{Ising}: Models without constraints that use spin variables (-1, 1).
\end{itemize}

\subsection{Adiabatic Quantum Computation}
\label{sec:quantumadiabatic}

In the constant pursuit of accelerating computational capabilities, quantum computing emerges as a solution to overcome the physical limits imposed by classical machines. In quantum computing, there are various models of computation.

The \textit{Adiabatic Quantum Computation} (AQC) model~\cite{farhi2000quantum} is capable of solving minimization problems on binary strings $\{x_1,...,x_n\}$ with an objective function $f : \{0,1\}^n \rightarrow R$. The variables of the problem represent the state of $n$ particles interacting with each other. A focal point of this model is the \textit{Hamiltonian} $H(t)$, which is a $2^n \times 2^n$ matrix, describing the possible energy levels at time $t \in [0, T]$. The eigenvectors of this matrix represent the system's eigenstates, while the eigenvalues represent their energies. The eigenstate with the minimum value is called \textit{ground state} $\phi(t)$.

An AQC algorithm has three fundamental phases:
\begin{itemize}
    \item The definition of an initial Hamiltonian $H(0)$, chosen so that its ground state $\phi(0)$ is easy to prepare;
    \item The definition of a final Hamiltonian $H(1)$, which describes the objective function, and its ground state $\phi(1)$ represents the optimal solution to the optimization problem.
    \item An adiabatic evolution path that controls the transition from $H(0)$ to $H(1)$, defined as:
        \begin{equation}
            H(s) = A(s)H(0) + B(s)H(1)
        \end{equation}
    where:
    \begin{itemize}
        \item $s = t/T$
        \item $A(s)$ is a function decreasing from 1 to 0;
        \item $B(s)$ is a function increasing from 0 to 1.
    \end{itemize}
\end{itemize}

According to the \textit{adiabatic theorem} by Max Born and Vladimir Fock~\cite{born1928beweis}, it is possible to achieve a transition that, if slow enough, allows the system to remain in the ground state throughout the process.

D-Wave machines do not strictly use an AQC model, which requires operating at zero temperature, but rather \textit{Quantum Annealing} (QA) devices~\cite{mcgeoch2013,Boixo2014}. QA is a more "messier" version that operates at positive temperatures, where qubits are strongly coupled to their environment while still maintaining some quantum coherence.

\section{Problem Statement} 
\label{sec:problem}

In a world with increasingly decentralized and cloud-based ICT, it becomes of vital importance to tackle cloud infrastructure issues such as resource sizing and scheduling, managing networking processes, managing and responding to cyber attacks, ensuring stable and fast services, and minimizing energy consumption. While the sought-after result for some of these is purely efficiency for better performance, energy consumption is a pivotal point to achieve not only a reduction in operating costs and, in some cases, performance improvement but also, or especially, a lower environmental impact.

\subsection{Topology and ILP Model}
\label{subsec:topology}
The development of an optimization model for energy consumption has become a key challenge in cloud infrastructure design~\cite{zheng2014joint, lu2018popcorns, jayanetti2019j-opt, amoretti2022classical}. Obtaining an optimal solution is possible for small instances, as the problem is NP-Hard~\cite{christensen2016multidimensional, hyafil1973graph, mr1976some}. For medium and large instances, the running time grows exponentially~\cite{amoretti2022classical}.

The energy efficiency of a \textit{Virtual Data Center} (VDC) is improved by reducing the energy consumption of \textit{Virtual Machines} (VMs) and networking using intelligent placement and routing techniques. Over the years, various models have been proposed. A first formulation as an \textit{Integer Linear Program} (ILP) was provided by Jin et al.~\cite{jin2013joint}, along with a unified representation method that converts the VM placement problem into a routing problem. They also introduced an approach that divides the network into clusters processed in parallel, along with a depth-first search-based algorithm.

Han et al.~\cite{han2015save} proposed two greedy bin packing heuristics for embedding algorithms to achieve near-optimal solutions, a path followed by other publications suggesting different types of heuristic algorithms~\cite{guo2020temperature,wang2021frequency,pham2017traffic,sun2017efficient,carpio2017replication}.

In~\cite{amoretti2022classical} and~\cite{colucci2023power}, two different topologies with their respective ILPs were proposed, later converted into a Quadratic Unconstrained Binary Optimization (QUBO) problem. This conversion enables quantum processing based on Quantum Annealing.

\begin{figure}[ht!]
    \centering
    \includegraphics[width=\columnwidth]{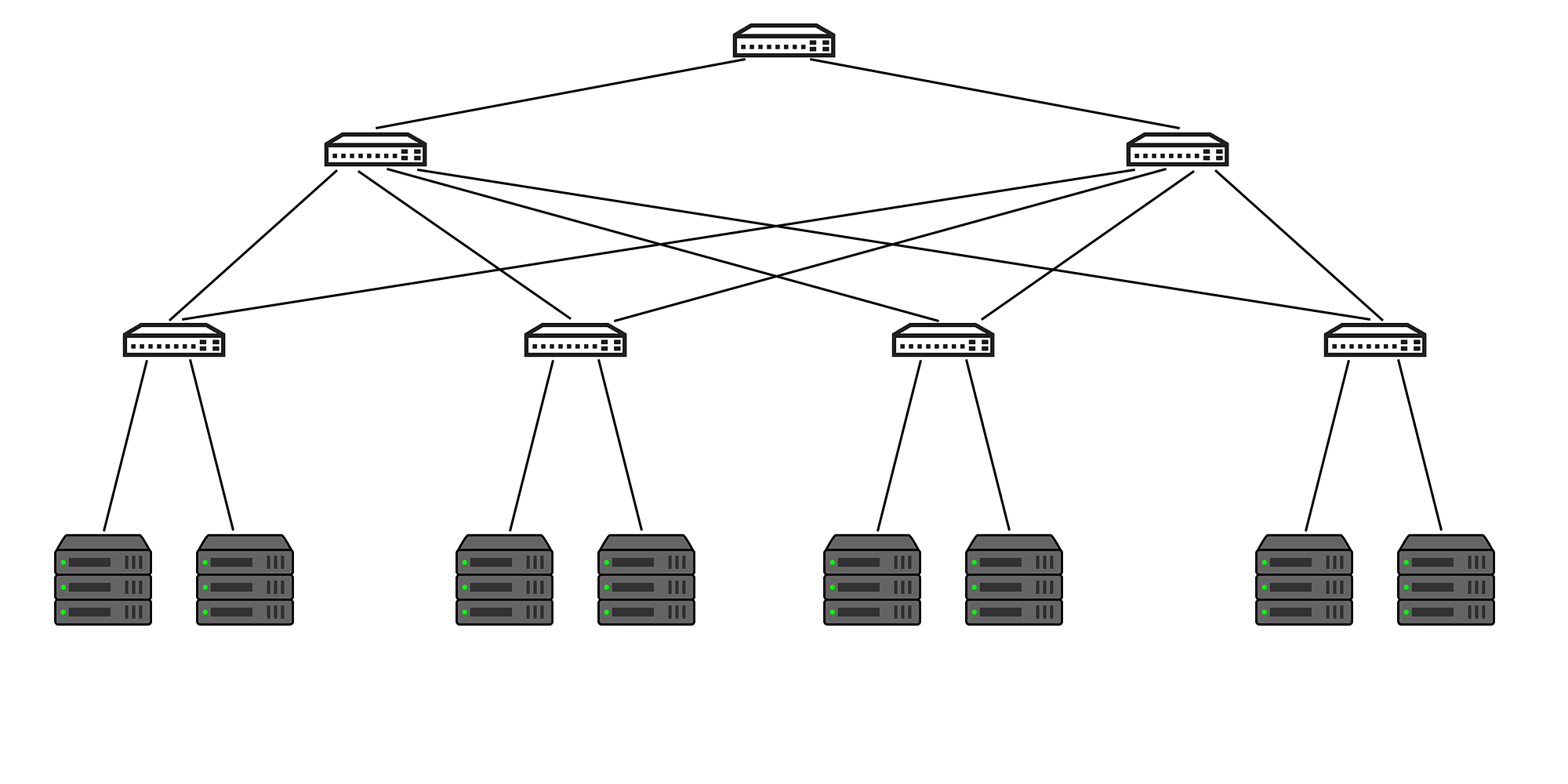}
    \caption{Servers and switches connected in a complex tree topology}
    \label{img:tree_topology}
\end{figure}

Because of its challenging complexity, which makes it a good benchmark, the ILP model defined in~\cite{amoretti2022classical} is considered in this work and further developed. The model concerns a physical system of servers connected via switches in a complex tree topology, where each switch is connected to all switches in the previous and next levels, ultimately reaching the last layer of switches before the servers, which is connected in a binary manner (Fig.~\ref{img:tree_topology}).

Despite the addition of connections between switches, the ILP model remains very similar to that proposed in~\cite{amoretti2022classical}. In particular, the structure is as follows (a quick summary of the notation is contained in Table~\ref{tab:notation}).

\begin{table}[ht!]
\begin{small}
    \centering
    \scalebox{0.8}{
    \begin{tabular}{|c|c|}
        \hline
        \textbf{Notation} & \textbf{Meaning} \\
        \hline
        $M$ & number of total servers \\
        \hline
        $N$ & number of total VMs \\
        \hline
        $K$ & number of total switches \\
        \hline
        $F$ & number of total flows \\
        \hline
        $s_{i}$ & test \\
        \hline
        $v_{ji}$ & $j$th VM on the $i$th server \\
        \hline
        $u(v_{ji})$ & the normalized CPU utilization of $v_{ji}$    \\
        \hline
        $sw_{k}$ & $k$th switch \\
        \hline
        $\rho(f, (n_{1},n_{2}))$ & part of the $f$th flow going from node $n_{1}$ to $n_{2}$ \\
        \hline
        $P(f_{n},( n_{1},n_{2}))$ & $\rho(f, (n_{1},n_{2}))+\rho(f, (n_{2},n_{1}))$ \\
        \hline
        $on(n_{1})$ & status of node $n_{1}$ \\
        \hline
        $on(n_{1},n_{2}) \:|\: l_{n_{1},n_{2}}$ & link between node $n_{1}$ and $n_{2}$ \\
        \hline
        $v_{src(f),i}$ & VM on $i$th server, transmitting on $f$th flow  \\
        \hline
        $v_{dst(f),i}$ & VM on $i$th server, receiving on $f$th flow \\
        \hline
        $P_{i}^{idle}$ & the idle power consumption of node $i$ \\
        \hline
        $P_{i}^{dyn}$ & dynamic power consumption of node $i$ \\
        \hline
        $C_{s}$ & capacity of server $s$ \\
        \hline
        $C_{l_{n_{1},n_{2}}}$  & capacity of the link between node $n_{1}$ and $n_{2}$  \\
        \hline
        $d_{f,l}$ & data rate of flow $f$ on link $l$ \\
        \hline        
    \end{tabular}
    }
    \caption{Notation for the problem formulation}
    \label{tab:notation}
\end{small}
\end{table}

\begin{small}
\begin{align}
    \begin{split}
        min &\sum_{i=1}^M \left( s_{i}P_{i}^{idle} \: + \: P_{i}^{dyn} \sum_{j=1}^{N}u(v_{ji})v_{ji} \right)  + \\
        &+ \sum_{k=1}^{K} \left( sw_{k}P_{k}^{idle} \: + \: P_{k}^{dyn} \sum_{f=1}^{F}\sum_{n \in M \cup K} (\: \rho(f, (n,k)) + \rho(f, (k,n)) \:) \right)
        \label{eq:objective}
    \end{split}
\end{align}
\end{small}

Subject to:

\begin{small}
\begin{align}
    &\sum_{j=1}^{N} u( v_{ji} ) v_{ji} \leq C_{i}s_{i}, \hspace{.65cm}\qquad \forall i \in \{1,..,M\}
    \label{eq:cons1}\\
    &\sum_{i=1}^{M} v_{ji} = 1  \hspace{1.95cm}\qquad \forall j \in \{1,..,N\}
    \label{eq:cons2}\\
    &\sum_{k=1}^{k} \rho(f,(i,k)) \leq v_{src(f),i},  \qquad \forall i \in \{1,..,M\}, \forall f \in \{1,..,F\}
    \label{eq:cons3}\\
    &\sum_{k=1}^{k} \rho(f,(k,i)) \leq v_{dst(f),i},  \qquad \forall i \in \{1,..,M\}, \forall f \in \{1,..,F\}
    \label{eq:cons4}\\
    \begin{split}
         &v_{src(f),i} - v_{dst(f),i} = \sum_{k=1}^{k} \rho(f,(i,k)) - \sum_{k=1}^{k} \rho(f,(k,i)),\\&\hspace{3.8cm} \forall i \in \{1,..,M\}, \forall f \in \{1,..,F\}    
    \end{split}
    \label{eq:cons5}\\[10pt]
    \begin{split}
         &\sum_{n \in M \cup K} \rho(f,(n,k)) = \sum_{n \in M \cup K} \rho(f,(k,n)),\\&\hspace{3.8cm} \forall f \in \{1,..,F\}, \forall k \in \{1,..,K\}
    \end{split}
    \label{eq:cons6}\\
    &\sum_{f=1}^{F}d_{f,l_{n_{1},n_{2}}} P(f_{n},( n_{1},n_{2})) \leq C_{l_{n_{1},n_{2}}} on(n_{1},n_{2}),  \qquad \forall l_{n_{1},n_{2}}
    \label{eq:cons7}\\
     &on(n_{1},n_{2}) \leq on(n_{1}),  \hspace{3.7cm} \forall l_{n_{1},n_{2}} 
    \label{eq:cons8}\\
    &on(n_{1},n_{2}) \leq on(n_{2}),  \hspace{3.7cm} \forall l_{n_{1},n_{2}} 
    \label{eq:cons9}
\end{align}
\end{small}

The objective (Eq.~\ref{eq:objective}) is divided into four fundamental summations:
\begin{itemize}
    \item the static power consumption of servers due to keeping them powered on to allocate VMs;
    \item the dynamic power consumption of VMs related to their performance requirements;
    \item the static power consumption of switches due to keeping them powered on to enable communications through them;
    \item the dynamic power consumption of switches related to incoming and outgoing flows connecting them to other nodes.
\end{itemize}

The variables of the problem are:
\begin{itemize}
    \item $s_{i}$, a binary variable indicating the status, on or off, of the \textit{i}th server.
    \item $v_{ji}$, a binary variable indicating the allocation of the \textit{j}th VM to the \textit{i}th server.
    \item $sw_{k}$, a binary variable  indicating the status of the \textit{k}th switch.
    \item $\rho(f, (n_{1}, n_{2}))$, a binary variable indicating if the \textit{f}th flow passes from node $n_{1}$ to node $ n_{2}$.
    \item $on(n_{1}, n_{2})$, a binary variable indicating whether the link between $n_{1}$ and $n_{2}$, with $n_{1} < n_{2}$, is on.
\end{itemize}

Eq.~\ref{eq:cons1} ensures that each VM can be assigned only to a powered-on server with enough capacity to host it.
Eq.~\ref{eq:cons2} guarantees that each VM is assigned to one and only one server.
Eq.~\ref{eq:cons3} ensures that outgoing flows from a server are less than or equal to the number of transmitting hosted VMs. The same applies to incoming flows and receiving VMs, guaranteed by Eq.~\ref{eq:cons4}. Additionally, it ensures that the difference between outgoing and incoming flows is exactly equal to the difference between the number of VMs assigned for transmission and reception, as guaranteed by Eq.~\ref{eq:cons5}.
The equality between outgoing and incoming flows in a switch is guaranteed by Eq.~\ref{eq:cons6}, while the correct balancing and assignment of flows, crucial to the routing problem, is ensured by Eq.~\ref{eq:cons7}.
Finally, Eq.~\ref{eq:cons8} and Eq.~\ref{eq:cons9} ensure that a link between nodes is on if and only if both are on and therefore necessary for routing a transmission.

The formulation methodology of this problem is crucial, as it is noticeable that it does not contain quadratic constraints nor non-binary variables. This has been done in anticipation of the need to convert it into a QUBO model to attempt direct resolutions using quantum computers.

Before delving into D-Wave solvers, it is essential to remind that the considered problem is formulated by means of a CQM, whose expression capabilities are superior to those of BQMs, which have access to only binary variables and no constraints at all.
Although the model has been designed to be convertible into a BQM, it does not mean that its conversion would be optimal nor efficient, even though necessary for computation on quantum machines.
Conversion between models is addressed in~\cite{hybridsolvers}, where it is noted that the transition between the various categories, both downstream (BQM to CQM) and upstream (CQM to BQM), is often suboptimal in the latter case, creating extremely complicated problems due to the relaxations necessary for adaptation.

\section{D-Wave Solvers} 
\label{sec:solvers}

\subsection{QPU Structure}
\label{subsec:solvers_QPU}
D-Wave computing systems rely on a \textit{Quantum Processing Unit} (QPU) that is essentially a complex arrangement of interconnected qubits, organized in a two-dimensional lattice denoted as Chimera~\cite{dwavesys}, to create a mapping network for optimization problems.

The qubits are microscopic rings of niobium, which exhibit quantum behaviors at low temperatures. The electric currents in these rings can circulate either clockwise or counterclockwise, determining the assignable spin value to the qubit (+1 or -1)~\cite{mcgeoch2013}. To ensure the quantum properties of niobium, each system is cooled to extremely low temperatures.

In the Chimera structure, qubits are oriented vertically or horizontally, intertwining in clusters, also known as unit cell. A component called coupler connects qubits of the same or different cluster. There are three types of couplers:
\begin{itemize}
    \item \textbf{Internal Coupler}: Connects each qubit to those in the same cluster but with orthogonal orientation (vertical-horizontal and vice versa).
    \item \textbf{External Coupler}: Connects colinear qubits, either vertically or horizontally aligned, but from different clusters.
    \item \textbf{Odd Coupler}: Connects qubits with the same orientation but not colinear.
\end{itemize}

During the processing, the system maps the QUBO (or Ising) problem onto the topology of the QPU, based on the objective function to be computed. 
Specifically:
\begin{itemize}
    \item each qubit represents a variable (node) of the objective function;
    \item each coupler represents a quadratic component (edge) of the objective function, specifically composed of the two variables (qubits) it connects.
\end{itemize}

Each of these components is then associated with the parameter (bias) defined in the objective function for calculating the final value.

Qubits have a limited and predefined number of connections, meaning that if the problem has edges between nodes not directly connected or an excessive number of edges on a node, it cannot be mapped directly. To address this issue, the system uses qubit chains, which are strong connections between qubits aiming to "merge" them into a single one, thus increasing the number of connections (and therefore neighbor qubits) it has access to. This is achieved through strong coupling by setting the value of the coupler that connects them to a negative value significant enough to ensure that they always have the same classical state at the end of the annealing phase, acting as a single variable. This process, called chaining, can be replicated with \textit{N} qubits, creating qubit chains long enough to satisfy any condition (within the limits of the developed architecture) \cite{dwavesysdoc}.

Although the quantum computing structure remains the same, D-Wave provides, through its SDK, two classes of solving methodologies for optimization problems~\cite{Ocean}, namely \textit{quantum solvers} and \textit{hybrid solvers}.
Their features are illustrated below.

\subsection{Quantum Solvers}
\label{subsec:quantum_solvers}
\begin{figure}[ht]
    \centering
    \includegraphics[width=\columnwidth]{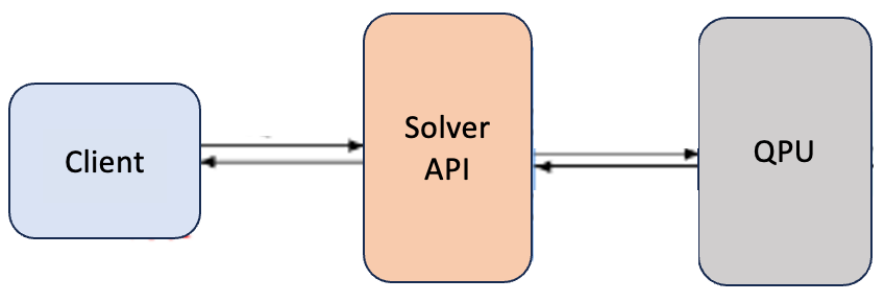}
    \caption{The full quantum methodology: the Solver API (front-end) is directly connected to the QPU (back-end).}
    \label{img:quantum_solver}
\end{figure}

The full quantum methodology~\cite{QuantumSolvers} works directly with the quantum machine without any intermediate steps, as seen in Figure~\ref{img:quantum_solver}, later undergoing the process of minor embedding for resolution. During this phase, the D-Wave machine manages the problem by attempting, through heuristic methods, to find the best possible embedding on its structure. Once found, a resolution algorithm, called Sampler, searches for a solution to the problem through a quantum annealing process.

Although this methodology is the only one to fully exploit the quantum technology, it is highly limited by the hardware constraints of the machine, not only by the number of qubits (and edges) within it, but also by the mapping, which may require qubit chains, further reducing the machine's capabilities, as observed in~\cite{colucci2023power}.

\subsection{Hybrid Solvers}
Using both classical and quantum resources to solve problems while maximising their complementary capabilities is known as quantum-classical hybridization. Ocean software currently supports two types of hybrid solvers: hybrid workflow solvers and Leap's hybrid solvers.

\subsubsection{Hybrid Workflow Solvers}
\begin{figure}[ht]
    \centering
    \includegraphics[width=\columnwidth]{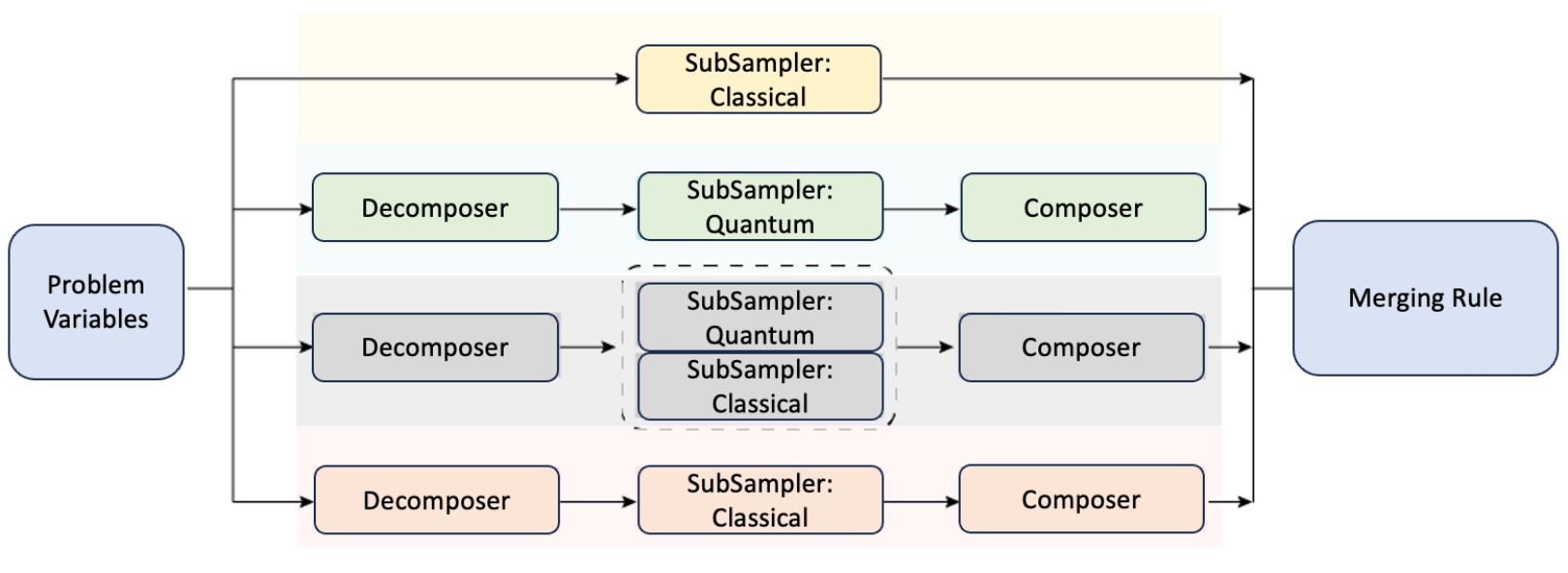}
    \caption{Example of hybrid workflow solver architecture.}
    \label{fig:hybridworkflow}
\end{figure}

The hybrid workflow methodology~\cite{HybridWorkflowSolvers} (illustrated in Fig.~\ref{fig:hybridworkflow}) aims to overcome the limitations of the full quantum methodology by processing the problem on multiple resolution flows (denoted as \textit{branches}), which are both quantum and classical. 
Each of these flows can contain a direct solver for the problem or a customized ensemble of the following components:
\begin{itemize}
    \item \textbf{Decomposer}: A decomposition algorithm that, given a complete problem and a decomposition rule, produces a solvable subproblem.
    \item \textbf{SubSampler}: A resolution algorithm, quantum or classical, for subproblems.
    \item \textbf{Composer}: A composition algorithm that, based on a defined rules, decides how to integrate the subproblem's result into the complete collection of variables (sample).
\end{itemize}

If multiple branches are present, a merging rule is required to determine how to fuse the various intermediate results.

\subsubsection{Leap's Hybrid Solvers}
\begin{figure}[ht]
    \centering
    \includegraphics[width=\columnwidth]{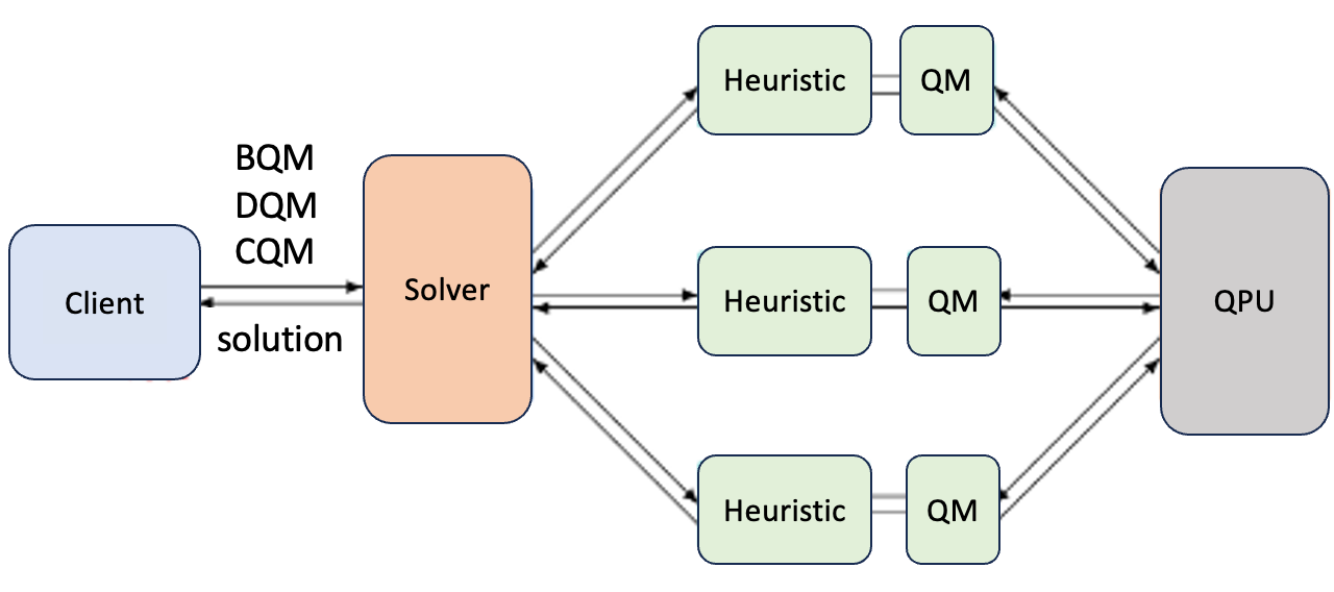}
    \caption{Generic flow structure of a Leap's hybrid solver.}
    \label{fig:Leap}
\end{figure}

The last methodology is that of Leap's hybrid solvers~\cite{LeapHybridSolvers,hybridsolvers}, which further explores the topic of hybrid computing, seeking to maximize its results. The general functioning of these systems, portrayed in Fig.~\ref{fig:Leap}, is based on a classical solver that processes the problem (either BQM, DQM, or CQM) and then starts one or more hybrid heuristic solvers that run in parallel to search for solutions. Each heuristic solver contains a classical \textit{heuristic model} that explores the solution space, and a unit called \textit{Quantum Module} (QM) that formulates quantum queries for the QPU. Replies from the QPU guide the heuristic module toward more promising regions of the search space, or to improve existing solutions.

\section{Results}
\label{sec:results}

The considered problem (illustrated in Section~\ref{sec:problem}) is formulated as a model with constraints, so even though it only presents binary variables, it is not directly processable by a BQM solver. The conversion led to a significant increase in computational difficulty and also made the BQM solver inefficient even at the early stages of tree depth (a problem also noted in~\cite{hybridsolvers}). On the other hand, the CQM solver, capable of processing the problem in its native state, was able to solve it optimally for various depths. It is crucial to point out that the number of variables (and constraints) increases exponentially as the tree grows, starting from a minimum of nearly 50 variables at tree depth 2 and increasing to almost $3 \cdot 10^4$ variables at tree depth 6, meaning that the unconstrained version becomes even more complicated after the relaxation.

For these reasons, only Leap's hybrid solvers produced correct results in reasonable times and only the CQM one was able to manage the problem through increasing tree depths, leading us to discard all the other methods and stick with the latter. The implementation and performance evaluation of the solution based on Leap's hybrid CQM solver are illustrated in the remainder of this section.

\subsection{Implementation}
To represent the tree structure, a class \texttt{Proxytree} was defined, containing both the tree structure (its depth, number of servers, adjancy list, number of switches, etc.) and its variables (servers capacities, switches capacities, flow capacities, etc.).
Once the tree is instantiated, a function creates the CQM problem using Ocean's methods. This is forwarded to another function which calls on D-Wave's cloud system Leap~\cite{Leap} for the problem resolution using the hybrid CQM solver. Solution value, times and other possible information is then saved.

To further evaluate this technologies, a split version of the problem was developed, which divides it in two phases. The first phase is the VM assignment on servers, the second one is the path planning between servers. Initially, this choice was made to the purpose of having a simpler problem to map on a QPU and permit a full quantum resolution. As soon as it became clear that only hybrid solvers were scalable enough for studying large problem instances,  the split version was used to evaluate how much the problem gets worsen by the suboptimal resolution.

The source code developed for this research work is available on GitHub\footnote{https://github.com/AmedeoB/quantum-optimization-performance}.

\subsection{Experimental Setup}
The tree creation values, except for the depth, remained the same for all the tests, with many being automatically scaled based on the selected tree dimension. The parameters are:
\begin{itemize}
    \item $C_s$ = 10, determines the servers capacities and the \textit{cpu\_util} of the VMs equal to 6 (choice made to grant that only one VM could be assigned to each server);
    \item $C_l$ = 5, which is then multiplied by the depth and decremented by 2 at each deeper level, determining the capacities of the links;
    \item $P^{idle}$ = 10, which is then multiplied by the depth and decremented by 5 at each deeper level, determining the idle power consumption of a tree node;
    \item $P^{dyn}$ = 2, which is then multiplied by the depth and decremented by 1 at each deeper level, determining the dynamic power consumption of a tree node;
    \item $d_{avg}$ = 4, average data rate that determines the data rate $d$ of all flows.
\end{itemize}

The lowering power consumption, both dynamic and idle, aims to encourage the solver to use higher level nodes only when necessary by giving them an higher cost. Communications between VMs are fixed and correspond to numbers ranging from 0 to $N-1$ paired in sequential order (0-1, 2-3, etc.). Also, the number of VMs is always equal to the number of servers.

By construction, the tree has always an even number of servers corresponding to a power of 2 determined by the depth. Increasing this parameter thus results in doubling the number of servers at each level, as well as a general doubling of the number of nodes.

During the tests with D-Wave's hybrid CQM solver, it emerged that the maximum tree depth for which a result could be obtained in both complete and split version is 6. Beyond this value, the system begins to struggle, no longer providing a result for the complete problem and, at the next level, not even for the path planning one. Moreover, already at depth 6, the problem proved to be unstable, failing to find a feasible solution for the complete problem 50\% of the runs. For this reason, the reported results stop at this tree depth.

It is important to note that, especially at greater depths, the hybrid CQM solver provides highly variable results, emphasizing the non-deterministic nature of the computing paradigm.

There are several classical libraries that deal with optimization using different techniques. It was decided to use DOcplex~\cite{docplex}, a Python modeling library developed by IBM using CPLEX~\cite{ilogcplex}.
Despite differences in function and parameter names, and in the data structures used for result management, the conversion from the D-Wave library~\cite{dwavesysdoc} is quite direct and linear, with variables, constraints, and objectives translating easily from one version to another.

DOcplex offers various stopping options, where the default is to find the optimal solution. Since the considered problem is complex and cannot rely on finding optimal solutions in acceptable times, two different stopping rules were decided.

The first stopping rule imposes a maximum computation time equal to the time taken by the hybrid system, to evaluate the efficiency of the result in equal response times. The second stopping rule imposes to return the first feasible solution found, to evaluate its efficiency, but above all, the time required to find it, which is crucial in comparing data at major tree depths.

Regarding the hardware used, the classical computation was performed on a machine equipped with 16 GB of RAM, an \texttt{AMD RYZEN 7 3800XT} processor and Windows 10 operating system, allocating 4 workers (cores) to computation.

Although the number of workers may seem limited, it was chosen after multiple tests involving both a number of workers from 1 to 16 on the specified configuration and a number of workers from 1 to 32 (which is the maximum number of workers that can be created) on a configuration equipped with an \texttt{INTEL XEON E5} and 50 GB of dedicated RAM. In all these configurations, it was observed that the growth of parallelization led to a slowdown, in some cases excessive, in computation speed due to overhead from the merging of parallel streams.

Even ILOG CPLEX documentation indicates that an increase in workers leads to a linear increase in memory consumption and that increased parallelization is not always the ideal approach, depending on the specific problem being tackled~\cite{ilogcplex}.

\subsection{Comparison Between Hybrid and Classical Solvers}

Before comparing hybrid and classical solvers, it is important to outline some peculiarities in the resolution of the problem at various tree depths, regardless of the adopted method. 

The resolution of the split version of the problem combines a trivial phase, VM assignment, to a non-trivial one, path planning. The peculiarity of this methodology, however, is the non-optimized assignment of VMs. In fact, having no particular constraints, the first part of the split problem makes decisions related only to the availability of servers, and not to the relative position of VMs in the network based on their connections. As shown in Table~\ref{tab:hybrid_energy_results}, at low tree depths the total energy values of the split problem are generally greater than the full model ones, which takes into account all the conditions.

\begin{table}[ht]
    \centering
    \scalebox{0.9}{
    \begin{tabular}{|c|c|c|}
        \hline
         ENERGY & & \\
        \hline
        \textbf{Tree Depth} & \textbf{Split Model} & \textbf{Full Model} \\
        \hline
        2 & 168.4 & 130 \\
        \hline
        3 & 453.6 & 376 \\
        \hline
        4 & 1185.6 & 1043.2 \\
        \hline
        5 & 2851.8 & 2921.6 \\
        \hline
        6 & 7178 & 54026.66 \\
        \hline              
    \end{tabular}
    }
    \caption{Table of total energy values for D-Wave's hybrid CQM solver at various tree depths.}
    \label{tab:hybrid_energy_results}
    
\end{table}

This advantage is also present in the classical version, denoting a fundamental peculiarity. As shown in Table~\ref{tab:calssical_energy_results}, the advantage of the complete problem lowers as the tree depth grows and, at depth 4, the split version performs better. This happens because the problem begins to be too complex for the solver, which benefits more from the simplified but suboptimal structure of the split problem, than from the optimal but complex structure of the complete one.

\begin{table}[ht]
    \centering    
    \scalebox{0.8}{
    \begin{tabular}{|c|c|c|c|c|}
        \hline
        ENERGY & \multicolumn{2}{c|}{\textbf{Time Limited}} & \multicolumn{2}{c|}{\textbf{Solution Limited}}\\
        \hline
        \textbf{Tree Depth} & \textbf{Split Model} & \textbf{Full Model} & \textbf{Split Model} & \textbf{Full Model} \\
        \hline
        2 & 178 & 130 & 178 & 178 \\
        \hline
        3 & 455 & 376 & 480 & 473 \\
        \hline
        4 & 1190 & 984 & 1652 & 2230 \\
        \hline
        5 & 2918 & 8025 & 3864 & 9312 \\
        \hline
        6 & NaN & NaN & 7291 & NaN \\
        \hline
    \end{tabular}
    }
    \caption{Table of total energy values for the CPLEX classical solver, time limited and solution limited, for various tree depths.}
    \label{tab:calssical_energy_results}
    
\end{table}

\subsubsection{Time Limited Solution}
The comparison method based on the first stopping rule (i.e., fixed execution timeframe) is the most direct and illuminating on the capabilities of the two solvers (D-Wave's hybrid CQM solver and CPLEX classical solver), confronting them on the individual quality of their results. 

\begin{table}[ht]
    \centering 
    \scalebox{0.8}{
    \begin{tabular}{|c|c|c|c|c|c|c|}
        \hline
        ENERGY & \multicolumn{2}{c|}{\textbf{Hybrid}} & \multicolumn{2}{c|}{\textbf{Classical}} \\
        \hline
        \textbf{Tree Depth} & \textbf{Split Model} & \textbf{Full Model} & \textbf{Split Model} & \textbf{Full Model} \\
        \hline
        2 & 168.4 & 130 & 178 & 130 \\
        \hline
        3 & 453.6 & 376 & 455 & 376 \\
        \hline
        4 & 1185.6 & 1043.2 & 1190 & 984 \\
        \hline
        5 & 2851.8 & 2921.6 & 2918 & 8025 \\
        \hline
        6 & 7178 & 54026.66 & NaN & NaN \\
        \hline
    \end{tabular}
    }
    \caption{Table of energy comparison between D-Wave's hybrid CQM solver and CPLEX classical solver with time limited computation, at various tree depths, and \% difference.}
    \label{tab:timelimited_results}
    
\end{table}

As shown in Table~\ref{tab:timelimited_results}, the energy of the solutions increases exponentially with the depth of the tree. Although it can be noted that hybrid solutions are generally better than classical ones, there are no significant differences in values up to a tree depth of 4. After this point, the energy of the classical resolution of the complete problem shoots up to quadruple the others. Finally, at depth 6, the classical solver fails to produce any feasible solution within the imposed time limit.

\subsubsection{First Feasible Solution}
By imposing to stop at the first feasible solution found, an attempt was made to explore in more detail how much time the classical solver took to produce a feasible result and, at the same time, to obtain results for the deeper levels that the classical solver failed to produce with limited time. 

With this method, it was sought to emphasize the weight of the advantage given by the QPU to the hybrid method, trying to identify how much the gap between methodologies amounted to once the critical point was reached.

\begin{table}[ht]
    \centering
    \scalebox{0.8}{
    \begin{tabular}{|c|c|c|c|c|c|c|}
        \hline
        ENERGY & \multicolumn{2}{c|}{\textbf{Hybrid}} & \multicolumn{2}{c|}{\textbf{Classical}} \\
        \hline
        \textbf{Tree Depth} & \textbf{Split Model} & \textbf{Full Model} & \textbf{Split Model} & \textbf{Full Model} \\
        \hline
        2 & 168.4 & 130 & 178 & 178 \\
        \hline
        3 & 453.6 & 376 & 480 & 473 \\
        \hline
        4 & 1185.6 & 1043.2 & 1652 & 2230 \\
        \hline
        5 & 2851.8 & 2921.6 & 3864 & 9312 \\
        \hline
        6 & 7178 & 54026.66 & 7291 & NaN \\
        \hline
    \end{tabular}
    }    
    \caption{Table of energy comparison between D-Wave's hybrid CQM solver and CPLEX classical solver with first solution limited computation, at various tree depths, and \% difference.}
    \label{tab:solutionlimited_energy_results}
\end{table}

\begin{table}[ht]
    \centering 
    \scalebox{0.8}{
    \begin{tabular}{|c|c|c|c|c|}
        \hline
        TIME [s] & \multicolumn{2}{c|}{\textbf{Hybrid}} & \multicolumn{2}{c|}{\textbf{Classical}}\\
        \hline
        \textbf{Tree Depth} & \textbf{Split Model} & \textbf{Full Model} & \textbf{Split Model} & \textbf{Full Model}\\
        \hline
        2 & 6.072 & 5.0438 & 0.0182 & 0.017 \\
        \hline
        3 & 10.186 & 5.261 & 0.057 & 0.044 \\
        \hline
        4 & 10.326 & 5.112 & 0.156 & 0.173 \\
        \hline
        5 & 15.758 & 9.107 & 0.662 & 5.261 \\
        \hline
        6 & 22.711 & 17.053 & 42.959 & 1501.65 \\
        \hline
    \end{tabular}
    }
    \caption{Table of time comparison between D-Wave's hybrid CQM solver and CPLEX classical solver with first solution limited computation, at various tree depths.}
    \label{tab:solutionlimited_time_results}
\end{table}

In Table~\ref{tab:solutionlimited_energy_results}, it is evident how the split solution of the classical solver is the less affected by the first-feasible-solution-found limitation, while the complete one visibly worsens compared to previous values. The energy gap between the solutions becomes more prominent at depth 4, especially between complete resolutions. Despite this, as shown in Table~\ref{tab:solutionlimited_time_results}, the classical solver always takes less than one second to produce these non-optimal but feasible solutions, and only at depth 5 we see a real increase in the minimum computation time to obtain an acceptable result. This not only requires a time similar to the hybrid CQM solver, but even produces an extremely worse solution. Moreover, the split model for the hybrid CQM solver takes more time than the full one. This highlights a clear lower bound, which is around 5 seconds, to problem resolution on D-Wave platform, caused by their automatic resolution time calculator applied to the solver.

A crucial point is the last depth level, where two aspects come to light.
The first is the solution of the split problem produced by the classical solver, which catches up to the hybrid CQM solver one, as shown in Table~\ref{tab:solutionlimited_energy_results}, but with a significant time gap of almost 90\%, as shown in Table~\ref{tab:solutionlimited_time_results}. This indicates how the support to the QPU guided the resolution of the problem, leading to a considerable speedup, as well as a minimal optimization of the energy value.

The second is the considerable increase in the gap between the split and complete solution, which not only underscores the marked difficulty that the problem is reaching but also highlights how the classical solver is unable, despite giving it almost 9000\% of the time used by the hybrid CQM solver, of producing a feasible solution, something that its competitor manages to do in less than 20 seconds.

\section{Conclusion} 
 
Despite D-Wave's QPUs alone are still too immature to handle excessively complex problems, the support of classical computers in hybrid systems leads to a significant performance increase compared to classical solutions. In summary, while these quantum machines represent a promising frontier for the future of mathematical optimization, at this stage their strength emerges primarily when synergistically integrated with classical computers.

Future work could deepen into D-Wave framework's limits, still using the same benchmark but trying different network topologies with different constraints and growth rates, to visualize a better break point for both quantum and hybrid approaches. Another possible development could be the design and implementation of completely different benchmarks, to see which of them are best suited for quantum solvers, and which of them are best suited for hybrid solvers.

\section*{Acknowledgement}
Michele Amoretti acknowledges financial support from the European Union – NextGenerationEU, PNRR MUR project PE0000023-NQSTI.
This research benefits from the HPC (High Performance Computing) facility of the University of Parma, Italy.

\bibliographystyle{IEEEtran}
\bibliography{main}

\end{document}